\documentclass[prl,twocolumn,showpacs,preprintnumbers,amsmath,amssymb]{revtex4}

\usepackage{graphicx}

\begin{document}

\title{Resistance Noise Scaling in a Dilute Two-Dimensional
Hole System in GaAs}

\author{R. Leturcq}
\affiliation{Service de Physique de l'Etat Condens{\' e}, CEA/DSM,
CE Saclay, F-91191 Gif-sur-Yvette, France.}
\author{D. L'H{\^ o}te}
\affiliation{Service de Physique de l'Etat Condens{\' e}, CEA/DSM,
CE Saclay, F-91191 Gif-sur-Yvette, France.}
\author{R. Tourbot}
\affiliation{Service de Physique de l'Etat Condens{\' e}, CEA/DSM,
CE Saclay, F-91191 Gif-sur-Yvette, France.}
\author{C. J. Mellor}
\affiliation{School of Physics and Astronomy, University of Nottingham,
University Park, Nottingham, NG7 2RD U.K.}
\author{M. Henini}
\affiliation{School of Physics and Astronomy, University of Nottingham,
University Park, Nottingham, NG7 2RD U.K.}

\date{\today }

\begin{abstract}
We have measured the resistance noise of a two-dimensional (2D)
hole system in a high mobility GaAs quantum well, around the 
2D metal-insulator
transition (MIT) at zero magnetic field. The normalized
noise power $S_R/R^2$ increases strongly when the hole density $p_s$
is decreased, increases slightly with temperature ($T$) at 
the largest densities, 
and decreases strongly with $T$ at
low $p_s$. The noise
scales with the resistance, $S_R/R^2 \sim R^{2.4}$, as
for a second order phase transition such as a
percolation transition. The
$p_s$ dependence of the conductivity is 
consistent with a critical behavior for such a transition, near a  
density $p^*$ which is lower than the observed MIT critical density $p_c$.

\end{abstract}


\maketitle

Two-dimensional (2D) dilute electronic systems at low temperature
offer the unique opportunity to study the physical effects of
strong Coulomb interactions. At low densities and in the limit of
weak disorder, the correlations between the carriers should
overcome the random motion of the electrons due to the fermions'
confinement. The relative magnitude of these two effects is
expressed by the ratio $r_s = E_{ee}/E_F$ between the Coulomb
interaction ($E_{ee}$) and the Fermi ($E_F$) energies, 
which is proportional to
$m^*/{p_s}^{1/2}$, $m^*$ being the effective mass of the carriers,
and $p_s$ their areal density. For $r_s \gg 1$, one expects to
observe a Wigner crystal \cite{Tan01}, thus raising the question of the
nature of the transition to this state by varying the density.
The recent observations of a metallic
behavior at intermediate $r_s$ values, $4 < r_s < 36$,
in 2D electron or hole systems (2DES or 2DHS) in high mobility
silicon metal-oxide-semiconductor field effect transistors
(Si-MOSFETs) and in certain heterostructures has
raised the possibility of a new metallic phase due to the
interactions \cite{Abr03}.
The metallic behavior is defined by a decrease of the resistivity
$\rho$ for decreasing temperature $T$, for $p_s$
larger than a critical density $p_c$.
In contrast, an insulating behavior ($d\rho/dT<0$) occurs for
$p_s<p_c$. However,
the nature of this metal-insulator transition (MIT) remains the
subject of ongoing debate \cite{Abr03,Alt06,Brun,Ham02,Lewal,Sara,
Vitka,Bog,Jar,Ilani,Gao01,Meir,Bene03,Spivak,HeShiperco,Shi01}.

Real systems are subject to disorder, which makes the physical
situation much richer. Experimentally,
around the MIT, the physical observables depend 
significantly on the disorder
\cite{Abr03,Ham02,Lewal,Sara,Vitka,Bog,Jar}.
Weak disorder could reduce the threshold for Wigner 
crystallisation $r_s^w$ from $r_s^w = 37 \pm 5$ \cite{Tan01}
to $r_s^w \approx 7.5$ \cite{Chui04}.
The system may also freeze into a glass \cite{Bog,Jar,Vak,Glass}
instead of crystallizing.
In GaAs 2DHS, recent local electrostatic studies \cite{Ilani} and
parallel magnetoresistance measurements \cite{Gao01} suggest
the coexistence of two phases.
Such a situation has been predicted by theories in which 
the disorder induces the spatial separation of a low and 
a high density phase \cite{Spivak,HeShiperco,Shi01}. In these models, 
the transport and other physical properties result from the
percolation of the most conducting phase through the 
insulating one. Such descriptions must be distinguished from
the percolating network of
Fermi liquid puddles connected by quantum point contacts \cite{Meir}: 
in this case, only one phase is present due to the absence of
interactions.

Resistance noise provides complementary
information to that obtained from DC transport experiments,
as shown recently in Si-MOSFETs close to the MIT
\cite{Bog,Jar}.
In this paper, we present the first resistance noise measurements
in a high mobility GaAs 2DHS at low densities, around the MIT. 
The density and temperature dependences of the normalized 
noise power $S_R/R^2$, indicate the onset of a new behavior as
the density is decreased.
We observe a clear scaling of the noise power vs. resistance,
suggesting the presence of a second order phase transition at a
density lower than $p_c$, which could be a percolation transition.
This possibility is supported by a scaling analysis of the
conductivity vs. density.

\begin{figure}
\includegraphics*[width=8.6cm]{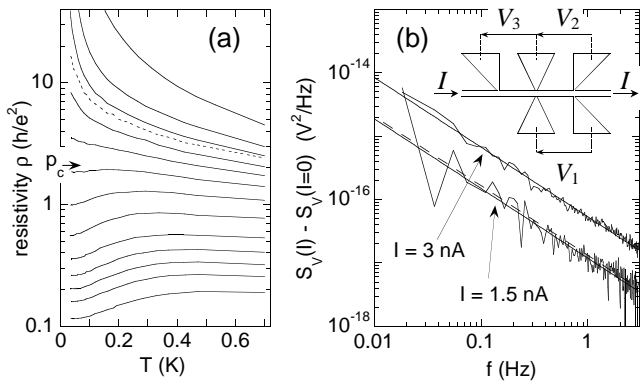}
\caption{(a) $\rho$ vs. $T$ for densities $p_s = $ 1.30, 1.39,
1.43, 1.44, 1.48, 1.52, 1.57, 1.63, 1.74, 1.86, 1.96, 2.06, 2.16,
2.31$\times 10^{10}$ cm$^{-2}$ (from top to bottom). The dashed line is the $\rho(T)$
curve corresponding to the $p_s$ limit ${p_c}' = 1.44 \times
10^{10}$ cm$^{-2}$, under which the activated law analysis 
is valid.
(b) $S_V$ vs. frequency at $p_s = 1.75 \times 10^{10}$ cm$^{-2}$ 
and $T=300$ mK,
for two currents $I = 3$ and $1.5$ nA. The continuous straight
lines are fits of the law $A/f^{\alpha}$ with $\alpha=1.1$, and
the dashed line is the fit for $I=3$ nA divided by 4.
Inset: Schematic view of the Hall bar, showing the three voltages
used in the correlation measurements.}
\label{fig1Bruit}
\end{figure}

The samples are 2DHS created in Si modulation doped (311)A GaAs
quantum wells. The front gate used to change the density is
evaporated onto a 1 $\mu$m thick polymide insulating film. This
reduces leakage between the gate and the 2DHS to a negligible
level and reduces the screening of the hole-hole interactions by
the gate. The mobility at a density $p_s=6 \times 10^{10}$
cm$^{-2}$ is $5.5 \times 10^5$ cm$^{2}$/Vs at 100 mK. The
experiments were carried out on Hall bars 50 $\mu$m wide, with a
distance of 300 $\mu$m between the voltage probes. The
fluctuations of the voltage were measured as a function of time,
for a fixed injected current $I$, low enough to ensure ohmic
conduction. The voltage noise power spectrum $S_V$ (obtained in
the $f=0.01-3$ Hz interval) is calculated by using
the cross-correlation technique \cite{Corr} for the two voltage noise signals
measured on opposite sides of the Hall bar ($V_{1}$ and $V_{2}$ in
the inset of Fig. \ref{fig1Bruit}(b)).
This allows us to minimize the noise due to the contacts and
preamplifiers (LI75A, NF Instruments). We verified that
$S_V(I)-S_V(0)$ was proportional to $I^2$ as expected for
resistance noise (see Fig. \ref{fig1Bruit}(b)), and we
define the resistance noise power as $S_R=(S_V(I)-S_V(0))/I^2$.
The possible contributions of external fluctuations
(e.g. gate voltage, current or temperature)
were ruled out by measuring the
correlation ${S_V}'$ between the voltage noise along two different
parts of the Hall bar ({$V_{1}$ and $V_{3}$ in the inset of Fig.
\ref{fig1Bruit}(b)). ${S_V}'(I)$ and ${S_V}'(0)$ were independent
of the frequency, contrary to $S_V(I)$, and their magnitudes were
identical, well below $S_V(I)$. We verified that a
correlation measurement on a single side of the Hall bar
gives the same noise magnitude as for the standard measurement, and
that $S_V$ measured on a 600 $\mu$m bar made of two contiguous
300 $\mu$m sections is close to
the sum of the values for each section;  
thus proving that possible
geometrical effects due to the finite width of the bar are weak.

Fig. \ref{fig1Bruit}(a) shows the 
temperature dependences of the resistivity $\rho$. 
The curves are similar to those obtained in
other high mobility 2DHS in GaAs
\cite{Abr03,Ham02}. The change of slope from $d\rho/dT>0$ to
$d\rho/dT<0$ attributed to the 2D MIT occurs at $p_c=(1.57 \pm
0.02) \times 10^{10}$ cm$^{-2}$ corresponding to $r_s^c \approx
24$ (assuming $m^*/m_e=0.37$ \cite{Hir02}). At densities below
${p_c}' = (1.44 \pm 0.05) \times 10^{10}$ cm$^{-2}$, an activated law
$\rho(T) \propto \exp(T_0/T)$ fits the data below a temperature
$T_l(p_s)$ which increases when $p_s$ decreases. Activated  
laws have been found in Si-MOSFETs \cite{Bog,Jar,Pudalov0,Shashkin}. 
$T_0$ depends linearly on
$p_s$ and vanishes at ${p_c}'$.

The resistance noise spectra have been measured for densities
$p_s$ ranging from $1.50$ to $1.78 \times 10^{10}$ cm$^{-2}$
(thus on both sides of the MIT at 
$p_c \simeq 1.57 \times 10^{10}$ cm$^{-2}$), and
for temperatures from 35 to 700 mK. They could be
satisfactorily fitted with a law $S_R(f)=A/f^{\alpha}$, $A$ and
$\alpha$ being the fit parameters. $\alpha$ does not show any
strong variation as a function of $p_s$ or $T$, and remains in the
interval $0.9-1.3$.
The normalized noise magnitude at 1 Hz, $S_R/R^2$(1\,Hz) = $A/R^2$ 
($R$ is the average resistance),
increases strongly when
$p_s$ decreases (see Fig. \ref{fig2Bruit}(a)), 
indicating the onset of a new regime at low
density. However, the continuous evolution as a function of $p_s$
could indicate that this regime is already present at $p_s > p_c$, as
suggested by local compressibility measurements
in $p-$GaAs \cite{Ilani} and noise measurements in Si-MOSFETs
\cite{Bog,Jar}. 
The onset of a new regime at low density also appears in the
temperature dependences: their slopes go from positive
to negative as $p_s$ decreases (see Fig. \ref{fig2Bruit}(b)).
A decrease of the noise magnitude when the temperature increases is 
expected for a degenerate system at low temperature \cite{FengetBir} or 
for localised electrons \cite{ShkloKogKoz}. The increase 
we observe at high density is thus of strong interest.
An increase is expected in the diffusive regime without 
quantum interference effects \cite{Wei01}, at high temperature.
In our case it could be related to the temperature dependent 
screening \cite{Senz} of the fluctuating disorder potential.

The temperature and density dependences of the noise
are qualitatively similar to those
in Si-MOSFETs \cite{Bog,Jar},
but we do not observe a comparable increase
of $\alpha$ from 1 to 1.8 when $p_s$ decreases.
The low level and different
nature of the disorder in $p$-GaAs in comparison to
Si-MOSFETs may contribute significantly to this difference.
Many studies have probed 
the physical importance
of the disorder magnitude 
\cite{Abr03,Ham02,Lewal,Sara,Vitka,Bog,Jar,Chui04},
but its nature could play a role too \cite{Spivak,Shi01,Gor01}.
The very short sample geometry in Ref. \cite{Bog} 
could also play a role, but similar behavior observed in much longer 
samples \cite{Jar} seems to speak against this possibility \cite{Note}.

\begin{figure}[b!]
\includegraphics*[width=8.5cm]{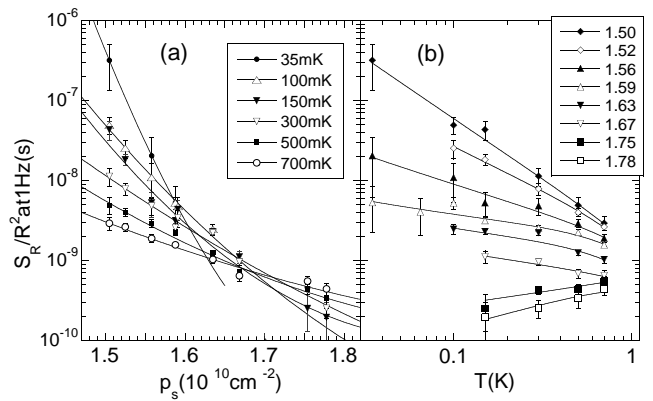} \caption{(a)
$S_R/R^2$(1 Hz) vs. density for six temperatures.
(b) $S_R/R^2$ vs. $T$, for eight
densities indicated in units of $10^{10}$ cm$^{-2}$. In (a) and (b), 
the lines are
guides to the eye.} \label{fig2Bruit}
\end{figure}

\begin{figure}
\includegraphics*[width=7cm]{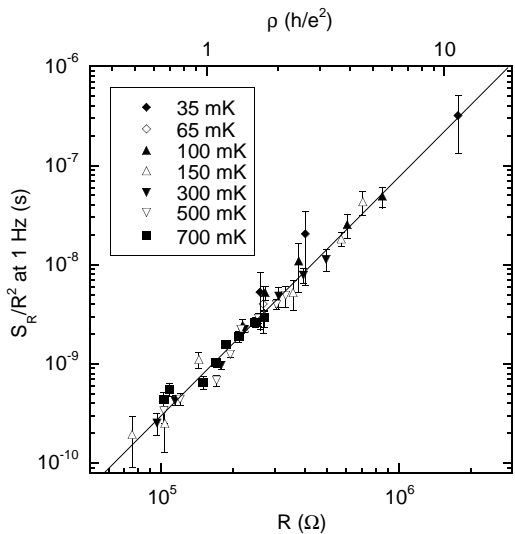}
\caption{Normalized noise power at 1 Hz as a function of the
resistance at various temperatures, and for all the densities
investigated between $1.5 \times 10^{10}$ cm$^{-2}$ and $1.78
\times 10^{10}$ cm$^{-2}$. The line is a fit with the power law
$S_R/R^2 \sim R^{2.4}$.} \label{fig3Bruit}
\end{figure}

Our main result is shown in Fig. \ref{fig3Bruit}. The normalized
resistance noise $S_R/R^2$ {\it scales as a function of the resistance}
$R$, whatever the density or the temperature, and {\it the dependence
is a power law}. A fit of all the points
gives $S_R/R^2 \sim R^{2.40 \pm 0.06}$. We verified that this
power law dependence remained true for each temperature.
This scaling suggests the existence of a second order phase
transition occuring at a critical density lower than $1.5\times
10^{10}$ cm$^{-2}$, the minimum density at which the noise has
been measured. It could be a percolation transition
as suggested by the theoretical arguments given earlier.

For a percolation transition, with $x$ the filling
parameter of the network and $x^*$ its critical
value, the conductivity $\sigma \sim 1/R$ vanishes, and 
$S_R/R^2$ diverges when $x \rightarrow x^*$ and $x > x^*$. 
They follow the scaling laws 
\cite{Kog04}
\begin{equation}
\sigma \sim (x - x^*)^{t} \qquad \text{and} \qquad S_R/R^2 \sim (x
- x^*)^{-\kappa} \quad , \label{scaling_perco}
\end{equation}
$\kappa$ and $t$ being respectively the critical exponents of the
resistance noise and conductivity. Thus,
\begin{equation}
S_R/R^2 \sim R^w \qquad \text{with} \qquad w =\kappa / t \quad .
\label{scal_noise_resist}
\end{equation}
As Eq. (2) is independent of $x$, it allows to 
investigate the scaling if $x$ is unknown. 
Our results would thus lead to $w=2.40 \pm 0.06$. 
Simulations of percolation transitions in 2D
yield $w$ = $1$ for a square lattice network
\cite{Ramm01}, and $w$ = $3.2$ for a continuous random-void model
\cite{Trem01}. Our case should be intermediate between
these limits, as the distance between the sites should be neither
constant, nor completely random.
Recent simulations of a 2D random resistance network yields
$w=2.6$ \cite{Penn02}. Such a scaling has been found in experiments 
on classical percolating 2D systems, with $w=3.4 - 4.2$ 
for sand blasted metal films \cite{Garf01}, and $w=2.0 \pm 0.1$ 
for thin gold films \cite{Koch01}.

We now consider the nature of the MIT we observe at $p_s=p_c$ 
in light of our results on the noise. Whether such a MIT is a
``true'' quantum phase transition has been discussed by many authors
\cite{Abr03,Alt06,Brun,Ham02,Lewal,Sara,Vitka,Bog,Jar,
Ilani,Gao01,Meir,Bene03,Spivak,HeShiperco,Shi01}.
A similar situation has been already studied in
three-dimensional systems, i.e.
the behavior of the resistance noise close to the
Anderson transition 
\cite{Coh1}. The noise was attributed to fluctuations of the 
system between the metal and the insulator due to the fluctuations 
of the disorder potential. The authors have shown theoretically and 
experimentally that such a system presents an exponential 
increase of the relative noise magnitude as a function of the 
resistance. In our case, the exponential dependence is incompatible 
with the data. We can thus draw the important conclusion that  
such a type of transition is excluded in our 2DHS.

\begin{figure}[b!]
\includegraphics*[width=8.6cm]{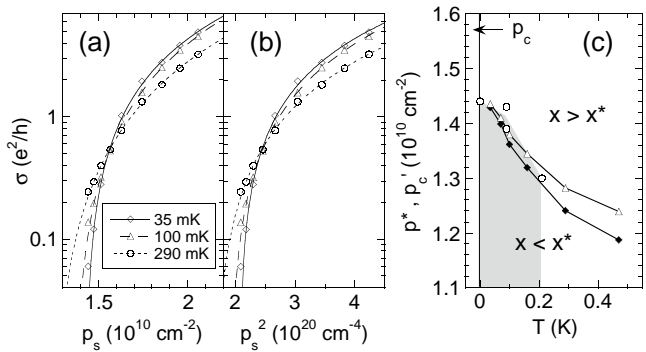} \caption{
(a) Conductivity vs. density for three temperatures. The lines are
fits with $\sigma \sim (p_s-p^*(T))^t$. (b) The same data as in
(a), plotted as a function of ${p_s}^2$. The lines are fits
with $\sigma \sim ({p_s}^2-{p^*(T)}^2)^t$. (c) Temperature
dependence of $p^*$. The closed diamonds (open triangles)
correspond to $\sigma \sim
(p_s-p^*)^t$ ($\sigma \sim ({p_s}^2-{p^*}^2)^t$).
Open circles: $p_s$ limit below which $\rho(T)$
can be fitted with an activation law $\propto
\exp(T_0/T)$. At $T=0$, this corresponds to $p_s={p_c}'$. The shaded
area is the domain where the activation law is valid.}
\label{fig4Bruit}
\end{figure}

To examine further the data in terms of a
percolation transition, we investigate the possible scaling of
the conductivity, i.e. $\sigma \sim (x - x^*)^{t}$.
The fact that our experimental points in Fig. \ref{fig3Bruit} scale 
together whatever the temperature suggests that $x$ is 
a function of both $p_s$ and $T$.
We consider the $x(p_s)$ dependence at a given low temperature.
The simplest 
relationship $x \sim p_s$ is obtained in models where the 
disordered potential
landscape is progressively filled by the Fermi liquid when the
density increases \cite{Meir}. Other models, in which the
interactions play the major role, predict the percolation of
a conducting phase in an insulating one \cite{HeShiperco,Shi01,Spivak}. 
In such models, $x$ is a more
complex function of $p_s$.
To investigate the first class of models we fitted the $\sigma$
vs. $p_s$ dependences for $p_s > 1.5 \times 10^{10}$ cm$^{-2}$
with the law $\sigma=\lambda (p_s -
p^*)^{t}$, $t$, $p^*$ and $\lambda$  being the parameters of the
fit.
The result is shown on Fig. \ref{fig4Bruit}(a) for three
temperatures. The fits are good, but the exponent $t$ varies
between 1.35 $\pm$ 0.1 and 1.9 $\pm$ 0.2 when the temperature is varied. 
The scaling
of the noise whatever the temperature suggests on the contrary
that $t$ is universal and independent of the temperature.
Moreover, a value close to 1.3 is expected in 2D systems
\cite{Meir}. 
We thus go to the second class of
models. Shi {\it et al.} suggested that $x \sim {p_s}^{\beta}$
with $\beta > 1$, in order to account for the experimental
compressibility measurements \cite{Shi01}. 
The fits of  $\sigma(p_s)$ with $\lambda' ({p_s}^\beta -
{p^*}^\beta)^{t}$ show that for $\beta \approx 2$, the $t$ values are
rather close to each other and to 1.3 when $T$ varies: they range
from 1.2 $\pm$ 0.1 to 1.45 $\pm$ 0.15.
Fig. \ref{fig4Bruit}(b)
shows the quality of the fits with $\sigma = \lambda' ({p_s}^2 -
{p^*}^2)^{t}$. The fact that $t$ has a weaker $T$
dependence and is closer to 1.3 thus favours the models in
which the interactions govern a phase separation.
The main conclusion of these $\sigma$ vs. $p_s$ studies
is that they are in agreement with a critical behavior of the conductivity,
thus {\it supporting} the 
assumption that the noise vs. resistance law is the signature of
a phase (percolation) transition.
As expected, the critical
densities $p^*$ extracted from the fits are lower than the minimum
$p_s$ value of the noise measurement range.

Fig. \ref{fig4Bruit}(c) gives the possible phase diagram 
of the percolation
transition, i.e. the temperature dependence of the critical
density $p^*(T)$ extracted from the fits of $\sigma(p_s)$. The
two lines correspond to $x \sim p_s$ (closed diamonds)
and to $x\sim p_s^2$ (open triangles). They are close
to each other and to the boundary (open circles)
between the two domains where the activation law analysis $\rho(T)
\propto exp(T_0/T)$ is valid or not, i.e. the $T_l(p_s)$
line mentioned above.
This result is consistent with the percolation interpretation:
below the percolation threshold, the system consists of isolated
conducting regions between which the conduction electrons jump due to
thermal activation. Such a description has been proposed to
analyse the properties of the insulator close to the quantum
Hall-insulator transition \cite{ShaetKuk}. 
The experimental result that the $B=0$
MIT is continuously connected to the
quantum Hall-insulator transition \cite{Han01} could thus be
related to their common percolating nature.

In conclusion, the resistance noise of a high 
mobility gated $p$-GaAs quantum well at low density
exhibits a huge increase of the noise power $S_R/R^2$ when
the density decreases, accompanied by a change of 
its temperature dependence slope.
The noise vs. resistance dependence
exhibits a scaling behavior for all densities and temperatures
studied, $S_R/R^2 \sim R^{2.4}$. The corresponding critical behavior
is confirmed by conductivity vs. density analyses which allow to 
extract the critical density $p^*(T)$ of a transition which 
could be of percolative nature. $p^*$ is lower than 
$p_c$, the usual MIT critical density. The non-exponential 
$S_R/R^2$ vs. $R$ dependence does not favor 
a ``true'' MIT at $p_s = p_c$.
The percolation could be that of 
a conducting phase in an insulating one as suggested by
theories of interacting electrons.

The authors thank J.-P. Bouchaud, G. Deville and D.C. Glattli
for useful discussions, and are grateful to
F. I. B. Williams for his help in initiating the project.


\begin{references}
\bibitem{Note}
The different behavior of $\alpha$ in our results compared to those of 
Refs [9,10] could also be due to our larger useful frequencies 
or to our smaller relative density interval.
\end{references}
\end{document}